\documentclass[aps,prl,twocolumn,superscriptaddress,nofootinbib,floatfix]{revtex4-1}
\usepackage{epsfig}
\usepackage{graphicx}
\usepackage{palatino}
\usepackage[english]{babel}
\usepackage{hyphenat}
\usepackage{amsmath}
\usepackage{amssymb}
\usepackage{mathtools}
\usepackage{mathrsfs}
\usepackage{slashed}
\usepackage{epstopdf}
\usepackage{xcolor}
\usepackage{braket}
\usepackage{booktabs}
\definecolor{lcolor}{rgb}{0.,0.0,0.}
\definecolor{citcolor}{rgb}{0,0.,0.5}
\usepackage[breaklinks,colorlinks,urlcolor=blue,citecolor=blue,linkcolor=blue]{hyperref}
\usepackage{multirow}
\usepackage{ltablex}
\usepackage{soul}

\newcommand{\beq}{\begin{eqnarray}}
\newcommand{\eeq}{\end{eqnarray}}

\newcommand{\bem}{\begin{multline}}
\newcommand{\eem}{\end{multline}}
\newcommand{\beg}{\begin{gather}}
\newcommand{\eeg}{\end{gather}}

\newcommand{\nn}{\nonumber\\}

\newcommand{\ben}{\begin{eqnarray*}}
\newcommand{\een}{\end{eqnarray*}}

\newcommand{\eqn}[1]{Eq.~\eqref{#1}}

\newcommand{\secn}[1]{Section~1}
\newcommand{\appn}[1]{Appendix~1}

\long\def\comment#1{ }

\def\and{\quad\text{and}\quad}

\def\0{{\boldsymbol 0}}

\def\k{{\boldsymbol k}}

\def\0{{\boldsymbol 0}}
\def\x{{\boldsymbol x}}
\def\y{{\boldsymbol y}}

\begin{document}

\title{Tracking Energy Loss in Heavy Ion Collisions}
\author{Jo\~{a}o Barata}
\affiliation{Physics Department, Brookhaven National Laboratory, Upton, NY 11973, USA}
\author{Ian Moult}
\affiliation{Department of Physics, Yale University, New Haven, CT 06511}
\author{Jo\~{a}o M. Silva}
\affiliation{Laboratório de Instrumentação e Física Experimental de Partículas (LIP), Av. Prof. Gama Pinto, 2, 1649-003 Lisbon, Portugal}
\affiliation{Departamento de Física, Instituto Superior Técnico (IST), Universidade de Lisboa, Av. Rovisco Pais 1, 1049-001 Lisbon, Portugal}

\begin{abstract}
The main modification to high $p_t$ jets evolving in the quark gluon plasma is the depletion of their energy to the thermal background. Despite the prominent role played by energy loss effects in jet quenching, their theoretical description is still limited, 
being, to a large extent, driven by phenomenological considerations. As the field moves towards the measurement of multi-scale jet substructure observables, it is necessary to develop new field theoretic techniques to describe energy loss. 
In this letter, we highlight a formal similarity between energy loss, and the renormalization group based track function formalism for computing jet substructure observables on charged particles.
We introduce an energy loss function, $L_i(x)$, tracking the energy fraction $x$ flowing from an initial hard parton $i$ down to the measured hadrons, and show that its renormalization group evolution can be studied in perturbative QCD, and is related to that of the track functions.
The features of this new approach are illustrated in the computation of projected energy correlators, opening the window to a better theoretical treatment of these observables in heavy ion collisions.  
\end{abstract}

\maketitle

\textbf{Introduction:} 
One of the key experimental signatures of the production of the quark gluon plasma (QGP) in heavy ion collisions at RHIC and the LHC is the suppression of the jet yield compared to the $pp$ baseline~\cite{PHENIX:2001hpc,ATLAS:2018gwx}. Such a modification is mainly driven by the quenching of the jets' energy to the thermal medium~\cite{Bjorken:1982tu}. Jet energy loss has received ample attention in jet quenching theory, leading to a substantial progress in the understanding and description of the main physical mechanisms driving this effect, see e.g.~\cite{Feal:2018sml,Barata:2021wuf, Caron-Huot:2010qjx,Sievert:2019cwq,Isaksen:2022pkj,Blaizot:2013hx,Mehtar-Tani:2017ypq, Mehtar-Tani:2017web,Apolinario:2014csa,Kuzmin:2024smy}.

Recent years have seen tremendous advances in the ability to study the substructure of jets in heavy ion collisions, see~\cite{Larkoski:2017jix,Asquith:2018igt,Apolinario:2022vzg} for reviews. These observables allow one to go beyond single high $p_T$ hadrons or jets as probes, to study correlations in energy flux, and hold the promise to drastically improve our understanding of the microscopic structure of the QGP. However, the theoretical description of these new classes of observables requires the development of techniques to account for the effect of energy loss of the entire parton shower.

The framework commonly applied to describe jet energy loss is still largely based on the early seminal work of Baier, Dokshitzer, Mueller and Schiff~\cite{Baier:2001yt}. In summary, their approach is based on the observation that the medium modified partonic cross-sections can be related to the vacuum ones via the simple relation: 
\begin{align}\label{eq:main_Qweights}
  \frac{d\sigma_i^{\rm med}}{dp_t} = \int_0^\infty d\varepsilon P_i(\varepsilon)     \frac{d\sigma_i^{\rm vac}}{dp'_t}\Big\vert_{p_t' = p_t+\varepsilon}\, .
\end{align}
Here $P_i(\varepsilon)$ is understood to be the \textit{probability for a parton $i$ with initial energy $p_t$ to lose energy $\varepsilon\ll p_t$ to the environment}. Thus, the energy shift of the vacuum cross-section, present on the right hand side of Eq.~\eqref{eq:main_Qweights}, allows to efficiently capture the experimentally observed shift of the final particle distributions, see for example~\cite{Apolinario:2024apr,Brewer:2018dfs,Casalderrey-Solana:2007knd,Mehtar-Tani:2013pia,ATLAS:2023iad,Milhano:2015mng,Rajagopal:2016uip}. Further, due to the steeply falling nature of the partonic vacuum cross-sections with $p_t$, the relation in Eq.~\eqref{eq:main_Qweights} can be further simplified to the form~\cite{Baier:2001yt}
\begin{align}\label{eq:Q_weight_simplified}
     \frac{d\sigma_i^{\rm med}}{dp_t} =  Q_i(p_t)  \frac{d\sigma_i^{\rm vac}}{dp_t} \, ,
\end{align}
where $Q_i(p_t)= \int\,  d\varepsilon \, e^{-n \varepsilon/p_t } P_i(\varepsilon)$, with $n$ the spectral index of the partonic cross-section. 

The relations in Eq.~\eqref{eq:main_Qweights} and Eq.~\eqref{eq:Q_weight_simplified} provide a simple prescription to capture energy loss effects in jet quenching physics and they have seen ample phenomenological usage, with the quenching weight factor $Q_i(p_t)$ being computed in increasingly more sophisticated models~\cite{Salgado:2003gb,Caucal:2021cfb,Mehtar-Tani:2021fud,Barata:2023bhh} based on perturbative QCD considerations~\cite{Baier:2001yt,Mehtar-Tani:2017ypq}. Notwithstanding the phenomenological success of this framework, it would be desirable to have a prescription of jet energy loss where a better treatment of the medium modifications can be given in QCD, avoiding to rely on cross-section level definitions, as in Eq.~\eqref{eq:main_Qweights}.

The goal of this \textit{Letter} is to provide an alternative non-perturbative description of in-medium energy loss, which can be studied using perturbative QCD methods. Our main insight is that medium induced energy loss shares formal similarities with the non-linear renormalization group equations developed in the study of track functions (TFs) \cite{Chang:2013rca,Chang:2013iba}. Making use of recent developments in the track functions formalism~\cite{Lee:2023xzv,Lee:2023tkr,Jaarsma:2022kdd,Li:2021zcf,Lee:2023npz}, we argue that this framework can be used to provide a new treatment of medium induced jet energy loss, that enables the calculation of energy loss in jet substructure observables, such as energy correlators, in heavy ion collisions~\cite{Barata:2024nqo, Barata:2023zqg, Andres:2024ksi,Andres:2022ovj,Yang:2023dwc,Bossi:2024qho,Singh:2024vwb,Xing:2024yrb}.

\textbf{Energy Loss Functions:} The main result of this letter is the introduction of a new non-perturbative function, the energy loss function (ELF), which can be used to describe energy loss in heavy ion collisions. To understand its motivation, we begin by briefly reviewing the track function formalism. Track functions were introduced in \cite{Chang:2013rca} to compute jet observables on charged particles. More generally, they provide a formalism for computing jet observables on any subset, $R$, of particles specified by some particular property, which achieves a rigorous factorization between perturbative and non-perturbative physics.

\begin{figure}
    \centering
    \includegraphics[width=.45\textwidth]{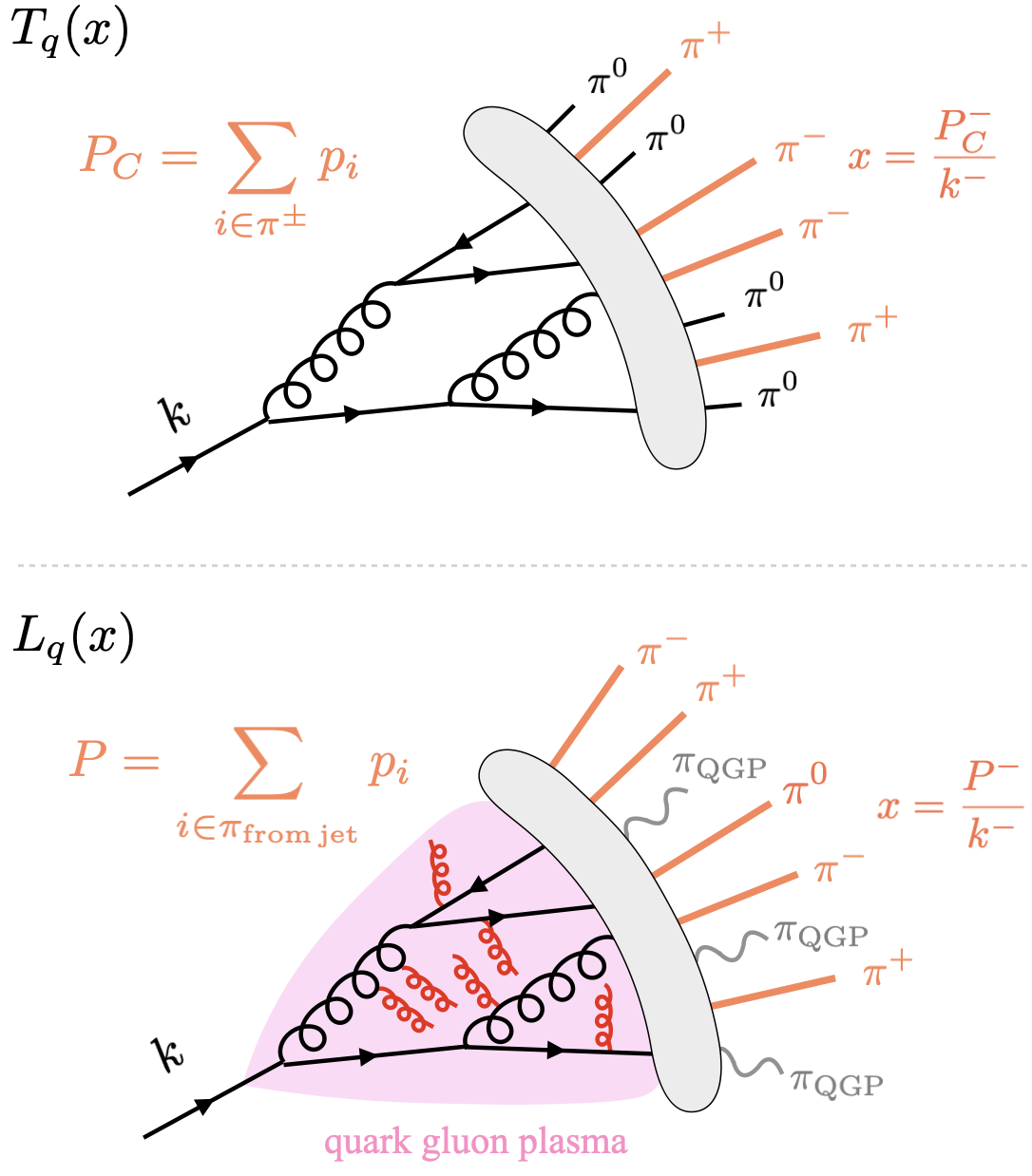}
    \caption{Illustration of the quark track (\textbf{top}) and energy loss (\textbf{bottom}) functions, in the limit where all particles hadronize to pions. For the track function, the energy fraction $x$ accounts only for the energy flow of charged hadrons. For the energy loss function, all hadrons coming from the jet contribute to the energy ratio, but pions coming from the thermal bath are not included.}
    \label{fig:ploit}
\end{figure}

Focusing for simplicity on the case of track functions for charged particles, we define the track function $T_i(x)$ as measuring the \textit{energy fraction $x$ carried by charged hadrons, that originated from an energetic parton $i$}, see Fig.~\ref{fig:ploit}. They can be given a field theory definition, which for the case of quarks, reads
\begin{align}\label{eq:def_tracks}
T_q(x) &= \int dy^+ d^2\y \, e^{i k^-\frac{y^+}{2}} \sum_{N,C} \delta\left(x- \frac{P^-_C}{k^-}\right) \frac{1}{2N_c}\nn 
&\times {\rm Tr} \left[ \frac{\gamma^-}{2} \langle \psi(y)|N,C\rangle \langle N,C|\bar \psi(0)\rangle\right] \, .   
\end{align}
Here we have split the state into a neutral component $N$, and a charged component $C$, and only the momentum of the charged component is measured. Gauge invariance is guaranteed by gauge links left implicit and a similar definition can be given for the gluon track function. As compared to fragmentation functions, which measure the energy fraction of a single hadron, track functions exhibit a more complicated non-linear renormalization group equation since they describe the energy fraction carried by \emph{all} hadrons of a particular property. Such equations have been extensively explored recently, and have been calculated at next-to-leading order  \cite{Jaarsma:2023ell,Chen:2022pdu,Chen:2022muj,Jaarsma:2022kdd,Li:2021zcf}, while the track functions have also been recently extracted from jet data~\cite{Lee:2023xzv,ATLAS:2024jrp}. In the heavy ion context, the modifications to the track functions induced by the medium have also been explored~\cite{Barata:2024nqo}. 

While using track functions in the calculation of physical observables is in general quite complex, they have recently become a practical tool for the specific case of energy correlator observables. Due to the fact that the energy correlators have recently been measured in heavy ion collisions \cite{CMS-PAS-HIN-23-004}, their theoretical simplicity makes them a primary target for theory development. For energy correlator observables, the only non-perturbative information that enters the calculation of the energy correlators are integer moments of the track functions  \cite{Chen:2020vvp}
\begin{align}
  T_i[n]  = \int_0^1 dx\, x^n T_i(n)\, ,  
\end{align}
where $T_i[0]=1$. Then, for example, the leading order EEC cumulative distribution on tracks \cite{Chen:2020vvp,Jaarsma:2023ell}, $\Sigma_{\rm trk}$, can be written as 
\begin{align}\label{eq:fact_theo}
    \frac{d\Sigma_{\rm trk}}{d \chi} = \sum_{ij} T_i[1]T_j[1]\frac{d\Sigma^{ij}_{\rm partons}}{d \chi} \, ,
\end{align}
where we summed over all partonic channels and $\chi>0$ denotes the separation angle between the two detectors used to define the EEC. This exhibits a clean factorization of the non-perturbative physics associated with the track measurement, from the partonic cross section.

Given the advances in the description of the renormalization group equations used for the study of track functions, we are motivated to apply these developments to the study of energy loss. In particular, one would like to achieve a formula analogous to \eqn{eq:fact_theo}, separating functions describing energy loss, from a vacuum cross section. We observe that the motivation for track functions, namely computing observables on a subset of particles, shares many similarities with energy loss. When considering energy loss, we are interested in studying energetic  jet constituents scattering off the medium, leading to the production of induced soft radiation, which results in an energy flow from the jet to the medium. Ultimately, this effect decreases the energy fraction carried by the hadrons coming from the partonic cascade, with part of the energy now carried by the hadronic remnants of the medium. 

Therefore, we are lead to define the energy loss function $L_i(x)$ which denotes \textit{the probability that an initial parton $i$, with energy $p_t$, goes to hadrons carrying $xp_t$ of the total momentum}; in the vacuum, it follows that $L_i(x)=\delta(1-x)$ to all orders, see Fig.~\ref{fig:ploit}. This recasting of energy loss in the language of track functions is the primary observation of this paper. Such a definition is sufficient for phenomenological applications, where a practical separation between bulk and jet constituents has to be implemented; a formal construction of this function in a framework that factorizes the collinear dynamics of particles within the jet from the dynamics of medium is beyond the scope of the current manuscript, and we leave it for future work.

The ELFs, as the TFs, are intrinsically non-perturbative objects. However, much like track functions, we can study their renormalization group evolution (RG) perturbatively. As should be clear from their definition, the ELFs exhibit the same evolution equations as TFs. In the context of heavy ion phenomenology, it is sufficient to consider the leading order RG evolution, which reads~\cite{Chang:2013iba,Waalewijn:2012sv}
\begin{align}\label{eq:RG_D}
  \mu \frac{d}{d\mu }L_i(x)  &= \frac{\alpha_s}{2\pi } \int dx_1 dx_2 dz\,  L_j(x_1) L_k(x_2) P_{i\to jk}(z) \nn 
  &\times \delta(x- (z x_1+(1-z)x_2)) \, ,
\end{align}
where $P_{i\to jk}$ denotes the leading order regularized QCD splitting functions. Although the TFs also satisfy Eq.~\eqref{eq:RG_D}, in the ELF context this RG evolution is only valid within the phase space region where the vacuum anomalous dimensions are unmodified by the medium; i.e. when $\mu$ is larger than the characteristic emergent medium scales.\footnote{We shall not discuss the particular shape of this boundary, and only note that, in practice, in the past such an effect has been introduced by imposing a phase space constraint in the integration in Eq.~\eqref{eq:RG_D}, see e.g.~\cite{Mehtar-Tani:2017web}.} Further, such scales, when appearing above the non-perturbative region, allow to directly compute the ELF, in analogy to what happens in the TF context for heavy quarks.

To illustrate this point and without making explicit use of a medium model, we assume that the presence of the background generates a (de)coherence scale $m_c$~\cite{Casalderrey-Solana:2012evi,Casalderrey-Solana:2011ule,Mehtar-Tani:2012mfa}, related to the ability of the medium to destroy quantum mechanical coherence in the parton shower due to multiple gluon exchanges with the QGP's quasi-particles. This scale regulates the fragmentation pattern, such that branchings occurring at angles smaller than $m_c/p_t$ are power suppressed and do not create additional color sources, i.e. $m_c$ plays the role of a quark mass. Thus, in analogy to boosted heavy quark effective theory (bHQET)~\cite{Fickinger:2016rfd,Beisert:2003jj,Bigi:1993ex,Benzke:2010js,Beneke:2023nmj},\footnote{We thank R. Szafron for first pointing out this analogy and to K. Lee for quantifying this observation.} and assuming the hierarchy of scales $p_t\gg m_c \gg \Lambda_{\rm QCD}\sim T\sim gT$, with $T$ the plasma temperature, one can further factorize the track functions into a perturbative track function and a non-perturbative element.\footnote{In the bHQET context this would involve a combination of shape functions~\cite{Fickinger:2016rfd} describing the hadronization of heavy quarks~\cite{Neubert:1993um}.} The details of this construction will be presented in a companion paper~\cite{New_heavyquark_EEC}. Here we highlight that the perturbative elements can be straightforwardly generalized from the bHQET context to jet energy loss, with the NLO partonic fragmentation function. While the energy loss effects, when neglecting hadronization effects,  can be modeled following known calculations for partonic energy loss~\cite{Mehtar-Tani:2017ypq, Baier:2001yt}, see Eq.~\eqref{eq:main_Qweights}. In what follows, we further comment on the implications of the presence of the scale $m_c$ when accounting for energy loss in jet observables.

 \textbf{Application to Jet Quenching:} To illustrate some of the properties of the ELFs, we exemplify now how they can be used in several aspects of jet quenching phenomenology. First, taking the first moment of Eq.~\eqref{eq:RG_D}
\begin{align}
\mu \frac{d}{d\mu }\langle x_i(\mu)\rangle  &= \frac{\alpha_s}{2\pi } \int dx_1 dx_2 dz\,  L_j(x_1) L_k(x_2) P_{i\to jk}(z) \nn 
  &\times (z x_1+(1-z)x_2) \, ,
\end{align}
one obtains the average energy loss, $ \langle \mathcal{E}_i \rangle =p_t- p_t \langle x_i \rangle$, evolution 
\begin{align}
   \mu\frac{d}{d\mu}\langle\mathcal{E}_i\rangle = 2\sum_{jk}(p_t - \langle\mathcal{E}_j\rangle)\gamma_2^{i\rightarrow jk} \, .
\end{align}
where $ \gamma_2^{i\rightarrow jk} = -\frac{\alpha_s}{2\pi}\int_z\, z  P_{i\rightarrow jk}(z)$. This procedure can be trivially extended to higher moments, which give access to energy loss fluctuations in the medium.

Beyond this single parton discussion, the ELFs also allow for a treatment jet energy loss in a systematic fashion. To that end we introduce the jet energy loss function $\mathbb{L}_i(x,p_t,R)$, which describes the \textit{probability for a jet of radius $R$, initiated by a parton $i$ with momentum $p_t$, to carry energy $xp_t$}. It can be related to the partonic energy loss functions as
\begin{align}\label{eq:big_L_i}
 \mathbb{L}_i(&x,p_t, R) = \frac{1}{2} \sum_{i} \int dx_1 dx_2 \, \mathcal{J}_{ij}(p_t, R,z,\mu) \nn 
 &\times L_{j}(x_1,\mu) L_{j}(x_2,\mu) \delta(x-(zx_1+(1-z)x_2)) \, ,
\end{align}
 where the leading matching coefficients, in vacuum, $\mathcal{J}_{ij}$ are given in~\cite{Waalewijn:2012sv}. The medium contributions can also be obtained, within a cut-off regularization scheme~\cite{Kang:2017frl,Li:2019dre}, and one finds, as an example for $i=j=q$, the following form
 \begin{align}
  \mathcal{J}_{qq}(p_t,R,z)= \frac{\alpha_s}{2\pi^2} \int_0^{p_tz(1-z)R} \frac{d^2\k}{\k^2} \Delta P_{q\to qg}(z,\k)\, ,
 \end{align}
 where $\Delta P_{q\to qg}(x,\k)$ denotes the (power) medium correction to the leading order regularized splitting function, which can be computed for several models of the underlying medium~\cite{Isaksen:2023nlr,Sievert:2019cwq}, which we shall not discuss here in detail. The generalization to other flavor channels is immediate. The jet energy loss function $\mathbb{L}_i$ can, in principle, also be constructed without matching to the ELFs, see e.g.~\cite{Lee:2023xzv}, but such an exercise is considerably more involved. 
 
 Using these elements, the probability that an energy fraction $x$ remains inside a jet initiated by a parton $i$ is given by
  \begin{align}
    \frac{1}{\sigma_i } \frac{d\sigma_i}{dx} = \frac{\mathbb{L}_i(x,p_t,R)}{2(2\pi)^3 J_i(p_t R)}\, ,
 \end{align}
where the normalization by $\sigma_i$ introduces the jet function on the right hand side, which can be obtained from the matching coefficients via the sum rule
 \begin{align}\label{eq:J_i}
   J_i(p_tR) = \int_0^1 dz \,  z\, \sum_j \mathcal{J}_{ij}(p_t,R,z)    \, ,
 \end{align}
 valid for both the vacuum and the medium components. The relations in Eqs.~\eqref{eq:big_L_i} -- \eqref{eq:J_i} provide the basic elements to describe the physics of energy loss in this framework, which can be systematically improved using standard perturbative techniques.

 \textbf{Application to Multipoint Energy Correlators:} The main application of the ELF formalism is to describe energy loss for observables that capture correlations between final state fluxes. Here we focus on the particular case of energy correlators \cite{Basham:1979gh,Basham:1978zq,Basham:1978bw,Basham:1977iq}. While energy correlators are an old observable, they have recently been made into a practical observable for jet substructure in hadron colliders \cite{Dixon:2019uzg,Chen:2020vvp,Komiske:2022enw}. They have now been measured in proton-proton \cite{Komiske:2022enw,CMS:2024mlf,Tamis:2023guc,ALICE:2024dfl}, pA, and AA collisions  \cite{CMS-PAS-HIN-23-004}. Due to their wide applicability in the heavy ion context, it is important to be able to properly describe energy loss effects for these observables.
 
 To that end we consider the simple (idealized) scenario of a $e^+e^-$ collision happening in the presence of a thermal background, which mimics the presence of the quark gluon plasma in heavy ion collisions. The projected energy correlator (PENC) cumulative distribution can be written as~\cite{Dixon:2019uzg,Lee:2022ige} 
\begin{align}
 \Sigma^{[n]}\left(\chi,\log \frac{Q^2}{\mu^2},\mu\right) &= \int_0^1 dx\, x^n J^{[n]}\left(\log \frac{\chi x^2 Q^2}{\mu^2},\mu \right) \nn 
 &\cdot H\left(x,\log \frac{Q^2}{\mu^2},\mu\right) \, , 
\end{align}
with the dot product taken in flavor space, and $Q$ the large momentum scale in the process. We consider here the leading order hard function $H=2(\delta(1-x),0)$, so that the PENC only depends on the jet function, which for the quark channel reads~\cite{Chen:2020vvp}
\begin{align}
   J_q^{[n]} = \frac{L_q[n]}{2^n} + \frac{\alpha_s}{4\pi}(-C_F) J_q^{[n],1} + \mathcal{O}(\alpha_s^2)\, .
\end{align}
The leading terms in $n$ at one-loop (setting the renormalization scale to $\mu^2=\chi Q^2$), while neglecting modifications to the jet function induced by the medium, can be written as
\begin{align}
  J_q^{[2],1}  &= \frac{37}{12} L_g[1]L_q[1]\, , \nn
  J_q^{[3],1}  &= \frac{611}{1200} L_g[2]L_q[1] + \frac{541}{300} L_g[1]L_q[2] \, .
\end{align}
This result can be directly compared to the one obtained by applying the quenching weight approximation in Eq.~\eqref{eq:Q_weight_simplified}, for which we find the leading one-loop terms $ J_q^{Q,[2],1} = \frac{37}{12} Q_g Q_q$ and $J_q^{Q,[3],1}  = \frac{249}{48} Q_g Q_q $, differing from the loss function based calculation for $n>2$. This mismatch between the two prescriptions appears at higher order in the strong coupling expansion; in particular for terms proportional to $\alpha_s^k$, the two approaches do not agree for $k>n$.

 \textbf{Extension to Higher Twist:} Although we have focused on the leading twist ELFs throughout this \emph{Letter}, we wish to highlight how they can be systematically extended to higher twist and the relevance of such terms to describe jet energy loss. 
 
 To this point in the implementation of ELFs, we have neglected the presence of correlations in the energy loss and hadronization process, analogously to was done for the TFs~\cite{Waalewijn:2012sv,Chang:2013rca}. While in the TF context charge correlations should only become important at scales close to $\Lambda_{\rm QCD}$, this might not be the case for ELFs, since multiparticle states might lose energy as a net effective color charge~\cite{Casalderrey-Solana:2007knd}, especially at scales of the order or smaller than $m_c$. Thus, in this regime it is not possible to associate each outgoing parton to a ELF, but rather one should make use of a \textit{multiparton} ELF. Notice that assuming that $m_c\gg \Lambda_{\rm QCD}$, then these functions can still be studied within a perturbative picture.\footnote{Although the same studies can be performed for TFs, the respective charge correlation corrections are power suppressed. For this reason, little attention has been given to this topic.} In such a case, one needs to change the factorization formula in Eq.~\eqref{eq:fact_theo} to\footnote{Here we use TFs, but the same discussion should follow for ELF, up to the introduction of the $m_c$ scale.} 
\begin{align}
    \frac{d\Sigma_{\rm trk}}{d \chi} = \sum_{ij} \int dx_1 dx_2\, T_{ij}(x_1,x_2) x_1 x_2 \frac{d\Sigma^{ij}_{\rm partons}}{d \chi} \, ,
\end{align}
where $T_{ij}(x)=\int_{x_1} T_{ij}(x_1,x-x_1) $ describes the \textit{energy fraction $x$ carried by the charged hadrons that originated from the system of energetic partons $i$ and $j$ as a whole.} In terms of field operators, this double index track function would assume a form which generalizes that in Eq.~\eqref{eq:def_tracks}:
\begin{align}\label{eq:def_tracks_ij}
T_{q\bar q}(z) &= \int dx^+ d^2\x \,  dy^+ d^2\y \, e^{i k_1^-\frac{x^+}{2}} e^{i k_2^-\frac{y^+}{2}} \nn 
&\times \delta\left(z- \frac{P^-_C}{k_1^-+k_2^-}\right) 
\mathcal{C}(x,y) \, ,
\end{align}
where $\mathcal{C}(x,y)$ involves four quark field operators at two separate spacetime points, and $P^-_C$ is the energy carried by charged states. At scales $\mu \gg \Lambda_{\rm QCD}$ one expects that this object can be written as a convolution of single TFs, while at smaller scales one would expect it to behave as the standard TF of an effective parton with the overall quantum numbers of the system. Finally, besides their phenomenological application, one expects the RG evolution for this combined TFs to satisfy a rather complex (leading) RG equation of the form 
\begin{align}
\mu \frac{d T_{ij}}{d\mu} &=  \mathcal{P}_{ij\to klm} \otimes T_{kl} \otimes T_{m}  \, ,
\end{align}
which we leave for future studies, along with the exploration of the structure of the higher twist ELF/TFs.

 \textbf{Conclusions:} The increasing sophistication of jet substructure measurements in heavy ion collisions necessitates an improved description of energy loss, incorporating many-particle correlations. In this short \textit{Letter} we have presented a new non-perturbative approach to the treatment of energy loss in heavy ions collisions, using theoretical tools developed for the calculation of jet substructure observables on charged tracks. We illustrated both how this treatment is related to more traditional energy loss formalisms, and how it can be used in the computation of jet observables, such as the PENCs. After discussing the discrepancies to the quenching weight formalism, we provided the necessary steps to systematically extend the discussion to jet energy loss, a critical element for any phenomenological jet quenching study. 

The theoretical elegance of the ELF approach, and its relevance for jet quenching phenomenology, motivates the extraction of these objects from Monte-Carlo simulations and, ultimately, real data. Since the TFs and the ELFs are sensitive to the full fragmentation pattern in the medium, it would be extremely valuable to compare data/MC extractions to theoretical predictions -- deviations between these results would give a clear indication for the presence of important non-perturbative effects and/or perturbative corrections to the in-medium cascade. Indeed, one can access the moments of the ELFs and TFs from the measurement of PENCs, as illustrated above, giving further motivation for a better theoretical description of these objects.

The ELFs provide strong motivation to understand the behavior of \textit{higher-twist} operators, which incorporate the presence of correlations in the hadronization and energy loss dynamics. Ultimately, we believe that the ELF formalism will enable a proper field theoretic description of correlations in jet substructure, advancing the heavy ion program.

 \textbf{Acknowledgments:}
 We wish to thank J. Brewer, Y. Go, K. Lee, A. V. Sadofyev, and R. Szafron for helpful discussions. We are especially grateful to R. Szafron for pointing out the analogy between certain aspects of jet quenching and boosted HQET, and to K. Lee for clarifying related questions and for providing a quantitative strategy to compute the TFs/ELFs. This work is supported by the European Research Council project ERC-2018-ADG-835105 YoctoLHC; by Maria de Maeztu excellence unit grant CEX2023-001318-M and project PID2020-119632GB-I00 funded by MICIU/AEI/10.13039/501100011033; by ERDF/EU; and by Fundação para a Ciência e a Tecnologia (FCT, Portugal) under project CERN/FIS-PAR/0032/2021. It has received funding from Xunta de Galicia (CIGUS Network of Research Centres); and from the European Union’s Horizon 2020 research and innovation programme under grant agreement No. 824093. JB is supported by the United States Department of Energy under Grant Contract DESC0012704. JMS is supported by Fundação para a Ciência e Tecnologia (Portugal, FCT) under contract PRT/BD/152262/2021.

\bibliographystyle{apsrev4-1}
\bibliography{EEC_ref.bib}

\end{document}